\documentclass[12pt]{article}
\usepackage[T1]{fontenc}
\usepackage[ansinew]{inputenc}
\usepackage{amsmath}
\usepackage{amssymb}
\usepackage{slashed}
\usepackage{graphicx}
\usepackage{psfrag}

\newcommand{\be}{\begin{equation}}
\newcommand{\ee}{\end{equation}}
\newcommand{\beqa}{\begin{eqnarray}}
\newcommand{\eeqa}{\end{eqnarray}}
\newcommand{\bseq}{\begin{subequations}}
\newcommand{\eseq}{\end{subequations}}

\newcommand{\pd}{\partial}

\renewcommand\a{\alpha}

\newcommand{\di}{\mathrm d}
\def\d{\partial}

\title{
{\huge {\sc
Comment on `Strong coupling in extended Ho\v rava--Lifshitz gravity'}}\\
 \date{}}
\author{D. Blas,\!$^a$ O. Pujol\`as,\!$^b$ S. Sibiryakov,\!$^{a,c}$\vspace{.2cm}\\
\normalsize\llap{$^a$}
 \it FSB/ITP/LPPC,
 \'Ecole Polytechnique F\'ed\'erale de Lausanne,\\
 \normalsize\it CH-1015, Lausanne, Switzerland\\
 \normalsize\llap{$^b$}\it CERN, Theory Division, CH-1211 Geneva 23, Switzerland\\
\normalsize\llap{$^c$} \it Institute for Nuclear Research of the Russian Academy of Sciences, \\
      \normalsize \it  60th October Anniversary Prospect, 7a, 117312 Moscow, Russia}
\begin{document}
\maketitle
\begin{abstract}
We show that, contrary to the claim made in arXiv:0911.1299, the extended Ho\v rava gravity model proposed
in arXiv:0909.3525 does not suffer from a strong coupling problem.
By studying the observational constraints on the model
we determine the bounds on the scale of the
ultraviolet modification
for which the proposal yields a phenomenologically
viable, renormalizable and weakly coupled model of quantum gravity.
\end{abstract}

\vspace{-15cm}
\begin{flushright}
\large CERN-PH-TH/2009-240
\end{flushright}

\vspace{14cm}

Building on the seminal works by P. Ho\v rava
\cite{Horava:2008ih,Horava:2009uw},
we have  recently proposed a power-counting renormalizable model
for quantum gravity without Lorentz invariance \cite{Blas:2009qj}. Remarkably, the model is free of
the pathologies \cite{Charmousis:2009tc,Li:2009bg,Blas:2009yd,Blas:2009qj,Koyama:2009hc}
present in the original Ho\v rava's
 proposal  and associated
with the additional mode of the gravitational excitations.
This is achieved by providing the extra mode with a proper quadratic
action around smooth backgrounds. We have argued in
\cite{Blas:2009qj} that this property
together with power-counting renormalizability of the theory
ensures that the theory is weakly coupled all the way up to
trans-Planckian energies\footnote{To be weakly coupled at {\em all}
  energies the model must fulfill the additional requirement that its
  marginal couplings do not develop Landau poles under the
  renormalization group flow. In other words, the theory must possess
  a weakly coupled UV fixed point. The Landau poles, if any, appear at
  exponentially high energies and are irrelevant for the purposes of
  the present paper.}.
We also argued in \cite{Blas:2009qj} that with appropriate
choice of parameters the theory satisfies
phenomenological constraints, and demonstrated this explicitly for
the simplest tests provided by the gravitational potential between static sources (Newton's
law) and homogeneous cosmology (Friedmann equation).

The consistency of the model presented in \cite{Blas:2009qj} has been recently
questioned in  \cite{Papazoglou:2009fj}, where
it is claimed that the model suffers from the same kind of strong
coupling problem as the previous versions of Ho\v rava's proposal \cite{Blas:2009yd,Koyama:2009hc}.
The aim of the present
note is to show that this claim is unfounded.
We support our arguments by considering a toy model where the absence of strong coupling
is demonstrated both using the power counting and at the level of scattering amplitudes.
We also analyze in more detail the observational bounds
on the model
and determine the window in the parameter space compatible with phenomenology
and weak coupling.

\subsubsection*{Review of extended Ho\v rava gravity}

We start by describing briefly the model \cite{Blas:2009qj}.
We consider the Arnowitt--Deser--Misner (ADM) decomposition
for the metric,
\be
\label{ADM}
\di s^2=(N^2-N_i N^i) \di t^2-2N_i  \di x^i \di t-\gamma_{ij}\di x^i \di x^j\;.
\ee
This decomposition defines a foliation of space-time by 3-dimensional
space-like surfaces thus splitting the coordinates into ``space''
and ``time''. We follow  \cite{Horava:2009uw} and,
unlike the case of general relativity (GR),
consider this foliation structure
as physical. This means that the
 group of invariance of the theory is not the full group of 4-dimensional
diffeomorphisms, but only its subgroup consisting of
foliation-preserving transformations
\be
\label{symm}
{\bf x}\mapsto\tilde{\bf x}(t, {\bf x})~,~~~t\mapsto\tilde t(t)\;.
\ee
The action of the model is taken in the form\footnote{The 3-dimensional indexes
$i,j,\ldots$ are raised and lowered using
$\gamma_{ij}$, and covariant derivatives are associated to
$\gamma_{ij}$.}
\be
\label{ADMact}
S=\frac{M_P^2}{2}\int \di^3x \di t \sqrt{\gamma}\,N\,\big(K_{ij}K^{ij}-
\lambda K^2 -{\cal V}[\gamma_{ij},a_i]\big)\;,
\ee
where $M_P$ is the Planck mass; $K_{ij}$ is the extrinsic curvature tensor
\be
\label{extr}
 K_{ij}=\frac{1}{2N}\left(\dot\gamma_{ij}-\nabla_i N_j-\nabla_j N_i\right)\;,
\ee
with the trace $K$;
$\gamma$ is the determinant of the
spatial metric $\gamma_{ij}$; $\lambda$
is a dimensionless constant. The ``potential'' term ${\cal
  V}[\gamma_{ij},a_i]$ in (\ref{ADMact}) depends on the 3-dimensional
metric and the lapse $N$. The latter enters into the
potential through the combination
\be
\label{ai}
a_i\equiv \frac{\d_iN}{N}\;,
\ee
which is covariant under the symmetry (\ref{symm}).
The dependence of the potential term on the gradients of the lapse is
the key difference of the model (\ref{ADMact}) compared to the
original Ho\v rava's proposal \cite{Horava:2009uw}.

The
existence of the preferred foliation structure reflects the
non-relativistic nature of the model: space and time enter into the
theory on
different footings. This allows to introduce into the action
terms with higher derivatives in spatial directions which improve the
ultraviolet (UV) behavior of the graviton propagator \cite{Horava:2009uw}; at the same time
the theory remains second order in time derivatives thus avoiding
problems with unitarity. Out of the
previous family of actions (\ref{ADMact}), one can
construct power-counting
renormalizable theories by considering the scaling transformations \cite{Horava:2009uw}
\bseq
\label{scaling}
\begin{gather}
\label{scaling1}
{\bf x}\mapsto b^{-1}{\bf x}~,~~~t\mapsto b^{-3}t\;,\\
\label{scaling2}
N\mapsto N~,~~~N_i\mapsto b^{2}N_i~,~~~\gamma_{ij}\mapsto\gamma_{ij}\;.
\end{gather}
\eseq
Under this scaling, the kinetic part of the action (\ref{ADMact})
and the operators of dimension\footnote{We asign dimension $-1$ to the
space coordinate. Then the dimension of time is $-3$, dimensions of
the lapse and the $3$-dimensional metric are zero, etc.}
6 in ${\cal V}$ are left unchanged (they are
marginal)\footnote{This is true classically.
At the quantum level one expects the coefficients in front of
 marginal operators to acquire logarithmic running  under the renormalization
group flow.}.
Operators of lower dimensions in
${\cal V}$ are relevant
deformations.
According to the standard arguments, considering operators up to
dimension 6 in the potential gives rise to an action which is
perturbatively renormalizable. Explicitly, the allowed
potential term is
\be
\begin{split}
\label{potential}
{\cal V}=&-\xi R -\alpha\, a_ia^i\\
&+M_P^{-2}(A_1\Delta R+A_2R_{ij}R^{ij}
+A_3a_i\Delta a^i+
A_4(a_ia^i)^2+A_5 a_ia_jR^{ij}+\ldots)\\
&+M_P^{-4}(B_1\Delta^2 R+B_2R_{ij}R^{jk}R_k^i+B_3a_i\Delta^2 a^i+
B_4(a_ia^i)^3+B_5 a_ia^ia_ja_kR^{jk}+\ldots)\;,
\end{split}
\ee
where $R_{ij}$, $R$ are the Ricci tensor and the scalar
curvature constructed out of the
metric $\gamma_{ij}$; $\Delta\equiv\gamma^{ij}\nabla_i\nabla_j$, and
$\xi$, $A_n$, $B_n$ are constants.
The
ellipses represent other possible  operators
of dimension 4 and 6
which can be constructed out of the metric $\gamma_{ij}$ and
are invariant under 3-dimensional diffeormorphisms\footnote{
Note that the operators with odd dimensions are forbidden by spatial
parity. Similarly, the terms in the action with one time derivative of
$a_i$ are excluded by the time-reversal invariance.
We stress
  that apart from these restrictions one must consider {\em all}
  operators of dimension up to 6 and compatible with the symmetries (\ref{symm})
to obtain a renormalizable action. Such operators are numerous and
only a few of them are written explicitly in the above
expression. The complete list of terms providing non-equivalent
contributions {\em at the quadratic level} is given in
\cite{Blas:2009qj}.}.
In what follows we set $\xi=1$, which can always be achieved by a
suitable
rescaling of time.
At low energies the potential is dominated by the operators
of the lowest dimension, namely,
the terms in the first line of (\ref{potential}). This leads
to the recovery in the infrared
of the relativistic scaling dimension $-1$ for both space
and time.

The
explicit breaking of 4-dimensional diffeomorphisms down to the
subgroup (\ref{symm}) gives rise to the presence of a new scalar gravitational
degree of freedom  \cite{Horava:2009uw,Blas:2009qj,Blas:2009yd}. Its
properties  at the
quadratic level  were analyzed in \cite{Blas:2009qj} where it was
shown that  the new mode is free of pathologies at all energies
(it is neither a ghost
nor a tachyon) in a wide range of parameters. The
proper behavior of the mode at low energies is ensured by the following
choice of the parameters $\lambda$ and $\alpha$ (see
Eqs.~(\ref{ADMact}), (\ref{potential}))
\be
\label{good}
0<\frac{\lambda-1}{3\lambda-1}~,~~~0<\alpha<2\;.
\ee
The additional mode does not have a mass gap:
at low energies it obeys a
linear dispersion relation with a velocity generically
different from that of gravitons (which is $1$ in our choice of
units). This signals the break down of Lorentz invariance
 down to arbitrary low energies.
 As we discuss
below, this has phenomenological consequences that
ultimately set bounds on the values of the parameters $\lambda$ and
$\alpha$ governing the low-energy physics of the model.

It is convenient to introduce
the covariant form of the theory which we obtain using the method described in \cite{Blas:2009yd}.
One encodes the foliation structure of space-time
 into a new St\"uckelberg field $\phi(t,{\bf x})$ by identifying  the surfaces of the foliation
with surfaces of constant $\phi$,
\be
\label{surf}
\phi(t,{\bf x})=const\;.
\ee
The invariance of the theory under reparameterization of
the foliation surfaces translates into the invariance under
reparamaterizations of $\phi$,
\be
\label{repar0}
\phi \mapsto f(\phi)\;,
\ee
where $f$ is an arbitrary monotonous function.
The quantities appearing in the action (\ref{ADMact})
reduce to the standard geometrical objects (induced metric,
extrinsic and intrinsic
curvature) characterizing the embedding of the hypersurfaces
defined by (\ref{surf})
in space-time. The central object in the
construction of these quantities is the unit normal
vector\footnote{The Greek indices
  $\mu,\nu,\ldots$ are raised and lowered using
the 4-dimensional metric $g_{\mu\nu}$, and
the covariant derivatives with these indices
are understood accordingly.}
$u_\mu$. Explicitly,
\be
\label{u}
u_\mu \equiv \frac{\nabla_\mu\phi}{\sqrt{\nabla_\nu\phi\,\nabla^\nu\phi} }~.
\ee
Note that $u_\mu$ is automatically invariant under the transformations
(\ref{repar0}). Other geometrical quantities associated to the
foliation are constructed out of $u_\mu$ and its derivatives, see
\cite{Blas:2009yd} for details. In this way
one obtains the following covariant
form of the action (\ref{ADMact}),
\be
\label{covar}
\begin{split}
S=-\frac{M_P^2}{2}\int \di^4x\sqrt{-g}\Big\{{}^{(4)}R
+(\lambda-1)(\nabla_\mu u^\mu)^2+\alpha\,u^\mu u^\nu
\nabla_\mu u^\rho\nabla_\nu u_\rho&\\
+\mathrm{(terms~with~higher~derivatives)}&\Big\}\;.
\end{split}
\ee
This action describes a scalar-tensor theory of gravity
invariant under 4-dimensional diffeomorphisms and the symmetry (\ref{repar0}).
Furthermore, after fixing the gauge $\phi=t$, it is equivalent to the
non-covariant form (\ref{ADMact}). Thus the
covariant (St\"uckelberg) formalism makes the presence of the extra scalar degree of
freedom explicit.

The first line in (\ref{covar}) contains all the terms with up to two
derivatives acting on $u_\mu$ and the metric; it
describes the low-energy physics of the model. Note that this
low-energy action is similar to a special case of the
Einstein-aether theory (see \cite{Jacobson:2008aj} for a recent
review). The difference from the
general Einstein-aether theory is that in our case the vector $u_\mu$ is,
by its definition (\ref{u}), hypersurface-orthogonal; i.e. it is characterized
by a single scalar field\footnote{In comparison
with \cite{Jacobson:2008aj}, we have absorbed one of the
free parameters of  the most generic action in a redefinition  of time ($\xi=1$).}.
Besides the low-energy
part, the full action of the model also contains the terms with higher
derivatives which we do not write explicitly. These terms arise from
the second and third lines in the potential
(\ref{potential}). Their important effect is to modify the dispersion relation
of the modes at high energies,
\be\label{dr}
E^2 = c^2 p^2 + \frac{p^4}{M_{*\,A}^2} + \frac{p^6}{M_{*\,B}^4}~,
\ee
where $E$ and $p$ are the energy and momentum of the modes, and\footnote{
To be precise, the lower expressions holds in the ``decoupling limit''
when $\alpha,~|\lambda-1|\ll 1$.
See \cite{Blas:2009qj} for the exact expression.
}
\be
\label{cs2}
c^2=
\begin{cases}
1 &\text{for helicity-2 modes}\\
c_s^2\equiv\frac{\lambda-1}{\alpha} &\text{for scalar graviton}.
\end{cases}
\ee
For simplicity, we  assume $c_s\sim 1$ in what follows. In this case, one reads off the
scales suppressing the higher derivative terms from (\ref{ADMact}), (\ref{potential}):
\be\label{M*s0}
M_{*\,A}\;,~~ M_{*\,B} \; \sim \;
\begin{cases}
A_i^{-1/2}M_P,~~B_i^{-1/4}M_P &\text{for helicity-2 modes}\\
\sqrt{\alpha} A_i^{-1/2}M_P,~~\alpha^{1/4}  B_i^{-1/4}M_P
 &\text{for scalar graviton.}
\end{cases}
\ee
As we now discuss, the presence of these
higher-derivative terms is
crucial to make the theory weakly coupled and renormalizable in the UV.

\subsubsection*{Would-be strong coupling and its resolution}

In the covariant language, the issue raised in \cite{Papazoglou:2009fj},
 can be understood as follows. Let us expand the low-energy action of the model (first line in
(\ref{covar})),  around the background consisting of the
Minkowski metric and the St\"uckelberg field linearly depending on
time,
\be
g_{\mu\nu}(t,{\bf x})=\eta_{\mu\nu}+h_{\mu\nu}(t,{\bf x})~,~~~\phi(t,{\bf x})=t+\chi(t,{\bf x})\;.
\ee
The result has the schematic form
$$
S=M_P^2\int d^4x \Big[-h\Box h - \alpha (\partial_i\dot\chi)^2 + (\lambda-1) (\Delta\chi)^2 
+ (\lambda-1) \dot\chi (\Delta\chi)^2 + \dots\Big]\,,
$$
where, for the sake of the argument, we have written down only one of the interaction terms.
The quadratic part of the perturbed action
remains invariant under the
relativistic scaling
\bseq
\label{relscaling}
\begin{gather}
\label{relscaling1}
{\bf x}\mapsto b^{-1}{\bf x}~,~~~t\mapsto b^{-1}t\;,\\
\label{relscaling2}
h_{\mu\nu}\mapsto b\,h_{\mu\nu}~,~~~\chi\mapsto\chi\;.
\end{gather}
\eseq
The interaction terms for both fields have positive dimensions with respect to this
scaling.
This means that these interactions would become strong at a certain
scale $\Lambda$ {\em if no new physics appeared at a lower scale}.
Under the assumption (motivated by phenomenological bounds) $|\lambda-1|\sim \alpha\ll 1$, the
covariant formalism with the action (\ref{covar})
allows to readily identify
the scale $\Lambda$ as
\be
\label{Lambda}
\Lambda=\sqrt{|\lambda-1|}M_P \sim \sqrt{\alpha}M_P\;.
\ee
The scale (\ref{Lambda}) has been erroneously
interpreted in \cite{Papazoglou:2009fj} as the UV cutoff of the theory
where the perturbative description breaks down.
Actually, $\Lambda$ is only the {\em cutoff of the low-energy
 approximation}. In the model described above, the would-be strong coupling is
actually not present if the higher-derivative operators (which
change the scaling dimensions of the interactions) enter into the
game at energies lower than (\ref{Lambda}) (see the related discussion in
\cite{Horava:2009uw}). For this to happen the energy scale of UV
physics \eqref{M*s0}, which we
collectively denote by $M_*$, must be
smaller than $\Lambda$,\footnote{One may worry that the choice of $M_*$ (and $\Lambda$) parametrically 
below $M_P$ introduces a fine-tuning in the model. Let 
us emphasize that this is not the case: having $M_*$ well below $M_P$ 
is technically natural. From the point of view of the low-energy 
theory, the reason is that the cutoff is set by $M_*$, and not $M_P$. 
Thus, neither $M_P$ nor $M_*$ receive large corrections.}
\be\label{M*Lambda}
M_* \lesssim \sqrt{\alpha} M_P~.
\ee
Then, the power-counting analysis performed in the ADM frame
(see \eqref{scaling}) shows that
under the new scaling the interactions are at most marginal, meaning
that there is
no strong coupling at the scale $\Lambda$. We conclude that the
correct interpretation of the scale (\ref{Lambda}) in the model at
hand  is that of the
scale suppressing the higher-derivative 
operators.

Let us illustrate our point by a simple toy model.
Consider a scalar theory with  action
\be
\label{action}
S=\alpha M_P^2\int
\mathrm{d}^4 x\left\{\Big(\varphi+\sum_{n\geq 2}a_n\varphi^n\Big)\left[-\Box
+\frac{\Delta^3}{M_*^4}\right]\varphi\right\}\;,
\ee
where the dimensionless coupling constants $a_n$ are assumed to be
somewhat smaller than 1.
This scalar theory shares all the relevant
properties with our actual gravity theory.
At low momenta, $|\Delta|\ll M_*^2$, the higher derivative terms can
be neglected and one obtains the following low-energy action
\be
\label{lowact}
S_{low\,E}=-\alpha M_P^2\int
\mathrm{d}^4 x\Big\{\Big(\varphi+\sum_{n\geq 2}a_n\varphi^n\Big)\Box
\varphi\Big\}\;.
\ee
Clearly, the invariance of the quadratic part of the action with
respect to the relativistic scaling transfromations
(\ref{relscaling1}) sets the scaling dimension of $\varphi$ to be 1.
The action contains irrelevant interactions under this
scaling which na\"ively become strong at the scale
\be
\label{Ltoy}
\Lambda=\sqrt{\alpha} M_P\;.
\ee
However, this is not the case provided $M_*<\Lambda$. At
momenta above $M_*$ the quadratic action is dominated by the term with
the highest number of spatial derivatives,
\be
S_{high\,E}^{(2)}=\alpha M_P^2\int \mathrm{d}^4 x\left\{\varphi\left[-\pd_0^2
+\frac{\Delta^3}{M_*^4}\right]\varphi\right\}\;.
\ee
This is invariant under anisotropic scaling transformations
(\ref{scaling1}) with $\varphi$ having scaling dimension
zero. Consequently, all the interactions in the full action
(\ref{action}) become marginal at high energies, and the
relative
strength of the interaction terms with respect to the free part
is always small.

It is instructive to see explicitly how the terms with higher derivatives
prevent the theory from becoming strongly coupled in the language
of scattering amplitudes. A well-known manifestation of the breakdown
of  perturbation theory is the saturation of unitarity bounds by
tree-level amplitudes (see e.g. \cite{Cornwall:1974km}). From the low-energy form
of the action (\ref{action}) one would conclude that tree-level unitarity is violated
at the scale (\ref{Ltoy}) and that perturbation theory is no longer valid
at higher energies.
As we shall now discuss, this conclusion would be incorrect.
This is essentially due to the peculiar kinematics of
 theories with anisotropic scaling, summarized by the dispersion
relation (\ref{dr}), which makes the unitarity bound much milder at
high energies as compared to the relativistic case.

To be concrete, we consider the $s$-channel scattering of two
$\varphi$ quanta with energy $E_0$ in the center of mass frame 
(which we assume to coincide with the preferred frame).
The optical theorem yields
for the s-wave amplitude\footnote{
We stick to the `relativistic' normalization of the 1-particle states
$|{\bf p}\rangle \equiv \sqrt{2 {\mathcal E}(p)} a_{\bf p}^\dagger
|0\rangle$.
This choice leads to conventional expressions for the amplitudes at
low energies where the relativistic dispersion relation is 
recovered.} ${\mathcal M}(2\to 2)$,
\be\label{optTh}
\begin{split}
&2 \mathrm{Im}\, {\mathcal M}(2\to 2)
=\\
&\sum_n\left(\prod_{i=1}^n
\int \frac{\di^3 p_i}{(2\pi)^3}\frac{1}{2{\mathcal E}(p_i)}\right)
\left|{\mathcal M}(2\to n)\right|^2(2\pi)^{4}
\,\delta\left(2E_0-\sum{\mathcal E}(p_i)\right)
\delta^{(3)}\left(\sum{\bf p}_i\right),
\end{split}
\ee
where the sum runs over all possible final states. Note that this relation holds
for arbitrary dispersion relation
\be
\label{dr1}
E={\mathcal E}(p)\;.
\ee
Being a sum of positive numbers, the r.h.s
of \eqref{optTh} is larger than any of the summands.
In particular,
\be
\label{Mineq}
2 \mathrm{Im}\, {\mathcal M}(2\to 2)
\geq
\int \frac{\di^3 p}{(2\pi)^2}\frac{1}{4{\mathcal E}^2(p)}
\left|{\mathcal M}(2 \to 2 )\right|^2
\,\delta\left(2E_0-2{\mathcal E}(p)\right),
\ee
Performing the integrations and using
$\mathrm{Im}\,{\mathcal M}\leq |{\mathcal M}|$ one obtains the bound
on the absolute value of the amplitude,
\be
\label{bound}
|{\mathcal M}(2\to 2)|\leq \;
16\pi\;{\mathcal E}'(p_0) \frac{E_0^2}{p_0^2}~,
\ee
where $p_0$ is related to the energy $E_0$ of the incoming particles by the
dispersion relation (\ref{dr1}). This bound simplifies when the
dispersion relation is given by a power-law,
${\mathcal E}(p)=p^z/M_*^{z-1}$,
\be
\label{bound2}
|{\mathcal M}|\leq 16\pi z\,\left[E_0/M_*\right]^{3(z-1)/z}~.
\ee
For relativistic particles, $z=1$, this takes the familiar form
$|{\mathcal M}|\leq 16\pi$. However, for the case of anisotropic
scaling with $z>1$ the bound (\ref{bound2}) is less restrictive and
allows the power-law growth of the amplitude with energy.

Let us check that in the model (\ref{action}) the bound (\ref{bound}) is
satisfied.
The dispersion relation
${\mathcal E}(p)=p\,\sqrt{1+ (p/M_*)^4}$ interpolates between $z=1$ and $z=3$
at low and high energies respectively.
The leading contribution to the tree-level amplitude comes from the diagram
\begin{figure}[h]
 \centering
\includegraphics[width=.14\textwidth]{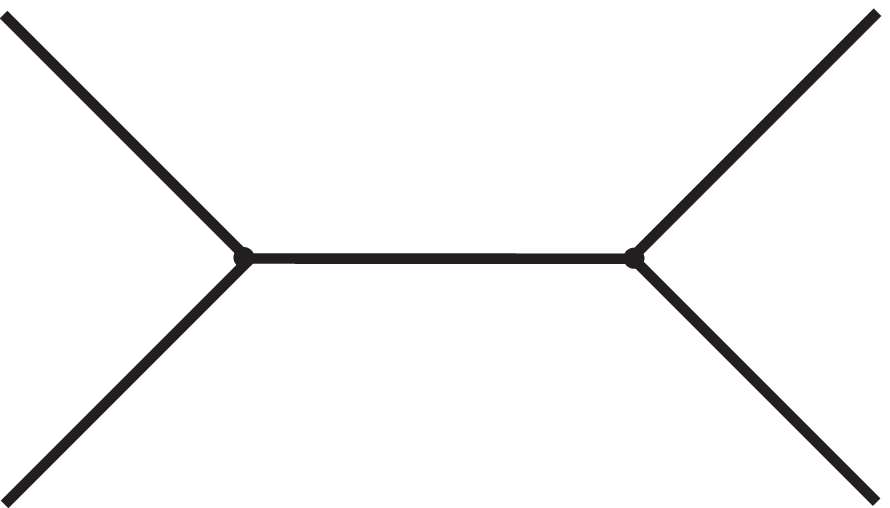}
\label{scatte}
\end{figure}\\
which is estimated as
\be
\label{amplit}
{\mathcal M}(2\to 2)\sim \frac{E_0^2}{\a \,M_P^2}\;,
\ee
where each vertex contributes a factor $E_0^2/\sqrt{\alpha} M_P$ 
and the propagator $1/E_0^2$.
At low energies, $p \ll M_* $, the bound (\ref{bound}) reduces to the
condition
$$
\frac{p^2}{\alpha M_P^2} \lesssim 1~.
$$
Na\" ively, this would imply the breakdown of tree-level unitarity
at  $\sqrt{\a}\, M_P$.
However, if $\sqrt{\a} M_P \gtrsim M_*$, the low-energy approximation fails
and the bound reads instead (for $p\gg M_*$)
\be
M_*^{2}\lesssim  \a  M_P^2,
\ee
which is indeed satisfied.
Thus, we recover the same result that was derived from
the scaling analysis (Eq (\ref{M*Lambda})): there are no signals that
perturbation theory is breaking down at the scale $\sqrt{\alpha}M_P$.

A few remarks are in order.
The above arguments do not exclude the
possibility that some marginal coupling of the theory develop a
Landau pole when loop corrections are taken into account.
If this turns out to be the case, 
the theory will become strongly coupled in the deep UV
(thus spoiling the UV-completeness of the proposal).
Even in this case, though, this would happen at an exponentially high energy.
For example, in the case of the toy model (\ref{action}) one can estimate this scale as
\[
\Lambda_{Landau}\sim M_*\exp\left[\frac{1}{\beta \;\big(a_n(M_*)\big)^\gamma}\right]\;,
\]
where $\beta$ and
$\gamma$ are numerical coefficients of order 1. 
Certainly, the presence or absence of Landau
poles in the extended Ho\v rava model (\ref{ADMact}) is an important
open issue requiring a
detailed renormalization group analysis of the theory.

Another basic issue concerning the
consistency of the theory  (\ref{ADMact}) at the quantum level
is to demonstrate the absence of anomalies in the
symmetry~(\ref{symm}).
We hope to return to these issues in the future.

\subsubsection*{Observational bounds on the UV scale}

The weak coupling condition (\ref{M*Lambda})  does not
allow to take the parameters $\alpha$, $|\lambda-1|$ to zero.
As already emphasized in \cite{Blas:2009qj}, this implies that
the model does not possess a GR limit in the IR, since inevitably
a gapless scalar polarization persists down to the lowest energies.
On the other hand, from the
remarkable success of GR in the description of low-energy gravitational
physics, one expects the
phenomenological constraints to put upper bounds on the parameters
$\alpha$ and $|\lambda-1|$, and hence on $M_*$. 
Thus the real physical
question is whether it is possible to
comply with observations without lowering the scale $\Lambda$
to an unacceptable level.

From the fact that the low-energy form of the action
(\ref{covar}) corresponds to a special case of the
Einstein-aether theory \cite{Jacobson:2008aj} one expects that
the phenomenology of the two models may be similar.
This expectation is supported by the results \cite{Blas:2009qj} for the
weak gravitational field of static
sources (at rest in the preferred frame) and for the expansion of the Universe.
Similarly to the case of Einstein-aether theory, static sources
in the model
(\ref{ADMact}) give rise to a linear metric which has the same form
as in GR with the Newton's constant\footnote{The expression (\ref{GN})
is obtained under the assumption that matter couples universally to
the metric $g_{\mu\nu}$. A more general situation compatible with
low-energy Lorentz invariance in the matter sector is
coupling it to the universal effective metric
$g^{eff}_{\mu\nu}=g_{\mu\nu}+\beta u_\mu u_\nu$, where $\beta$ is a
dimensionless parameter. This modification preserves the GR form of the
weak gravitational field
but changes the expression for the Newton's constant.}
\be
\label{GN}
G_N=\big(8\pi M_P^2(1-\alpha/2)\big)^{-1}\;.
\ee
Importantly,
this implies that the PPN parameter $\gamma^{PPN}$ has its GR value,
$\gamma^{PPN}=1$.
The cosmological expansion in the model (\ref{ADMact}) is governed
by the standard Friedmann equation with the effective gravitational
constant $G_{cosm}\neq G_N$, which again coincides  with the
situation in the Einstein-aether theory.
The
phenomenological constraint  $|G_{cosm}/G_N-1|\lesssim 0.13$
\cite{Carroll:2004ai} sets
a mild bound \cite{Blas:2009qj}: $\alpha$, $|\lambda-1|\lesssim 0.1$.

Following the guide of the Einstein-aether, one expects the most
stringent constraints on the model (\ref{ADMact}) to come from the
observational bound on the PPN parameter $\alpha_2^{PPN}$ which
characterizes the preferred frame effects due to Lorentz violation
(see \cite{Will:2005va}
for the precise definition)\footnote{The constraint due to the
  absence of gravitational Cherenkov emission by high-energy cosmic
  rays \cite{Elliott:2005va}
is easily evaded by setting the velocity of the scalar graviton
(as well as that of the helicity-2 mode)  larger or equal than the
maximal velocity of matter particles.}.
The detailed study of the PPN
corrections in the model (\ref{ADMact}) will be reported
elsewhere\footnote{The details of the PPN calculations in the model
  (\ref{ADMact}) are different from the Einstein-aether case due
  to the absence of the transverse vector mode.} \cite{pheno};
here we only sketch the estimate for $\alpha_2^{PPN}$.
This parameterizes an angular
dependent contribution to the Newtonian potential produced by a source
moving with velocity $v$ with respect to the preferred frame,
$$
\Phi_N = -\frac{G_Nm}{r}\left(1+\frac{\alpha_2^{PPN}}{2} \,v^2 \,\sin^2\theta\right)\;,
$$
where $m$ is the mass of the source, and $\theta$ 
is the angle between the radial vector and the velocity of the 
source with respect to the preferred frame, $\cos\theta = {\bf \hat r} \cdot{\bf  \hat v}$.
From the physical
point of view, this contribution is due to the interaction via the
Lorentz-violating scalar mode associated with the vector $u_\mu$
in the action (\ref{covar}) ({\em cf.} discussion in \cite{Cheng:2006us}). The result (\ref{GN})
for static gravitational field
shows that for $\alpha\ll 1$ the scalar-exchange amplitude  is suppressed by
$\alpha$.
Thus we conclude that\footnote{An explicit computation \cite{pheno}
  yields $\alpha_2^{PPN}=\alpha^2/2(\lambda-1)=\alpha/2c_s^2$, where
  $c_s$ is defined in (\ref{cs2}). This
  coincides with (\ref{estim}) when $c_s$ is of order one.}
\be
\label{estim}
\alpha_2^{PPN}\sim\alpha\;.
\ee
The bound on $\alpha_2^{PPN}$ following from the observed alignment of
the rotation axis of the Sun with the ecliptic  \cite{Will:2005va} gives
\[
 \alpha,~|\lambda-1|\lesssim  4\times 10^{-7}\;,
\]
where we again assume $\alpha$ and $|\lambda-1|$ to be comparable.
This translates into the bound on the suppression scale for
higher-derivative operators
\be
\label{Lambdapheno}
M_*\lesssim  10^{15}\, \mathrm{GeV}\;.
\ee
To our knowledge, this is the strongest upper bound arising from
gravitational physics.

The lower bound on $M_*$ from purely gravitational physics is very
mild. Direct tests of Newton's law at the distances $\sim 10\,\mu$m imply \cite{Will:2005va}
\be
\label{lowermild}
M_*\gtrsim 0.1 \, \mathrm{eV}\;.
\ee
A stronger bound may be obtained under the additional assumption that
$M_*$ also sets the suppression scale for terms with higher powers of
momentum $p$ in the dispersion relations of the matter fields,
specifically, of photons. Timing of
active galactic nuclei \cite{Albert:2007qk}
and gamma ray bursts
\cite{Collaborations:2009zq} constrains the value of such terms. Note
that odd powers of $p$ can be forbidden (at least, in
electrodynamics) by imposing parity. Then, the leading contribution to
the dispersion relation has the form $p^4/M_*^2$ which yields
\cite{Albert:2007qk,Collaborations:2009zq}
\be\label{lower}
M_* \gtrsim 10^{10} \div 10^{11} \mathrm{GeV}~.
\ee
Let us stress that, unlike the upper bound (\ref{Lambda}), this lower
bound is model dependent: it relies on the assumption that the
UV modification to the dispersion relation for photons appears at the same scale as that for
scalar graviton. This need not hold in some formulations of the theory.

It is worth emphasizing the difference between the situation in the model
(\ref{ADMact}) and that  in
Horava's original proposal, both in its projectable and non-projectable
versions \cite{Horava:2009uw}. In both versions,
the strong coupling scale calculated within the low-energy
effective theory is so low that the introduction of any new physics at
that scale is phenomenologically unacceptable. Indeed, as discussed in
\cite{Blas:2009yd}, in the non-projectable case the strong coupling
scale for the additional mode is inversely proportional to the curvature
radius of the background. It goes to zero for flat, cosmological and
static backgrounds, invalidating the proposal.

In the projectable case it was shown \cite{Blas:2009qj,Koyama:2009hc}
that the scalar graviton mode is unstable at large wavelengthes.
The requirement that the rate of the instability is smaller
than the age of the Univers (in order not to spoil standard cosmology)
gives the bound
$|\lambda-1|^{1/2}\lesssim H_0/M_*$, where
$H_0$ is the present value of the
Hubble parameter and $M_*$ is the suppression scale of the
higher-derivative operators
\cite{Blas:2009qj,Koyama:2009hc}. On the
other hand, $M_*$ must be smaller than
the strong coupling scale of the low-energy theory, which in this case
is \cite{Koyama:2009hc} $|\lambda-1|M_P$. This gives the bound
$M_*\lesssim (H_0^2M_P)^{1/3}\sim (1000 \, \mathrm{Km})^{-1}$.
Comparing this with the experimental bound (\ref{lowermild}), one
concludes that the projectable case in the weakly coupled regime is ruled out.
\\

To sum up, we have shown that the claim \cite{Papazoglou:2009fj} about
the presence of strong coupling problem in
the model (\ref{ADMact}) is unfounded. The absence of strong coupling
is actually a built-in
feature of the model. It suffices for the scale $M_*$ suppressing
the higher derivative operators to be slightly lower than the naive strong
coupling scale calculated in the low-energy theory.
The observational constraints on deviations of gravity
from GR place an upper bound $M_* \lesssim 10^{15} \mathrm{GeV}
$. Under the additional (model dependent)
assumption that this scale is common for gravity and matter sectors
one obtains a lower bound $M_* \gtrsim 10^{10} \div 10^{11} \mathrm{GeV}$.
Within this range, to the best of our knowledge, the model is compatible with the existing
data. Thus, so far, the model is a phenomenologically viable
candidate for a renormalizable quantum theory of gravity.
Needless to say, whether the theory is truly renormalizable (anomaly free) and UV complete
remains an important open issue. Another major question is the
mechanism for the
recovery of Lorentz invariance in the matter sector at low energies
(see \cite{Collins:2004bp,Iengo:2009ix} for the detailed discussion of
the problem).

\paragraph*{Acknowledgments}

We thank Gia Dvali, Sergei Dubovsky, Roberto Franceschini, Duccio
Pappadopulo and Alex Vikman
for useful discussions.
This work was supported in part by the Swiss Science Foundation
(D.B.), the Tomalla Foundation (S.S.), RFBR grants 08-02-00768-a and
09-01-12179 (S.S.)
and the Grant of the President of Russian Federation
NS-1616.2008.2 (S.S.). S.S. thanks Stanford Institute for Theoretical
Physics and Center for Cosmology and Particle Physics at NYU for
hospitality
during the work on this paper.

\end{document}